\documentclass[10pt, nodraftcls, twocolumn]{IEEEtran}  

\usepackage{amsmath,graphicx}
\usepackage{amssymb}
\usepackage{algpseudocode}
\usepackage{cite}
\usepackage{amsfonts}
\usepackage{graphicx,subfig}
\usepackage{fancyhdr}  %
\usepackage{cases}
\usepackage{extarrows}
\usepackage{algorithm}
\usepackage{multirow,tabularx}
\usepackage{mathtools}
\usepackage{xcolor}
\usepackage[english]{babel}

\usepackage{float}
\usepackage{subfig}
\usepackage{cases}

\usepackage{footnote}
\makesavenoteenv{tabular}
\makesavenoteenv{table}

\IEEEoverridecommandlockouts   

\usepackage{hyperref}
\hypersetup{%
  bookmarks=true
}

\begin{document}

\title{Holographic MIMO Surfaces for 6G Wireless Networks: Opportunities, Challenges, and Trends}

\author{
\IEEEauthorblockN{Chongwen~Huang,~\IEEEmembership{Member,~IEEE}, Sha Hu,~\IEEEmembership{Member,~IEEE}, George~C.~Alexandropoulos,~\IEEEmembership{Senior Member,~IEEE}, Alessio Zappone,~\IEEEmembership{Senior Member,~IEEE}, Chau Yuen,~\IEEEmembership{Senior Member,~IEEE}, Rui Zhang,~\IEEEmembership{Fellow,~IEEE,} Marco Di Renzo,~\IEEEmembership{Fellow,~IEEE,} and M\'{e}rouane Debbah,~\IEEEmembership{Fellow,~IEEE}}
\vspace{-0.5cm}
}
\maketitle


\begin{abstract}
Future wireless networks are expected to evolve towards an intelligent and software reconfigurable paradigm enabling ubiquitous communications between humans and mobile devices. They will be also capable of sensing, controlling, and optimizing the wireless environment to fulfill the visions of low-power, high-throughput, massively-connected, and low-latency communications. A key conceptual enabler that is recently gaining increasing popularity is the Holographic Multiple Input Multiple Output Surface (HMIMOS) that refers to a low-cost transformative wireless planar structure comprising of sub-wavelength metallic or dielectric scattering particles, which is capable of impacting electromagnetic waves according to desired objectives. In this article, we provide an overview of HMIMOS communications by introducing the available hardware architectures for reconfigurable such metasurfaces and their main characteristics, as well as highlighting the opportunities and key challenges in designing HMIMOS-enabled communications.
\end{abstract}

\section{Introduction}
Future wireless networks, namely beyond fifth Generation (5G) and sixth Generation (6G), are required to support massive numbers of end-users with increasingly demanding Spectral Efficiency (SE) and Energy Efficiency (EE) requirements \cite{Akyildiz2018mag,6Gvision,Marco2019,husha_LIS1}. In recent years, research in wireless communications has witnessed rising interests in massive Multiple Input Multiple Output (MIMO) systems, where Base Stations (BSs) are equipped with large antenna arrays, as a way to address the 5G throughput requirements. However, it is still a very challenging task to realize massive MIMO BSs with truly large-scale antenna arrays (i.e., with few hundreds or more antennas) mainly due to the high fabrication and operational costs, as well as due to the increased power consumption.

Future 6G wireless communication systems are expected to realize an intelligent and software reconfigurable paradigm, where all parts of device hardware will adapt to the changes of the wireless environment\cite{Akyildiz2018mag,Marco2019,LIS_twc2018}. Beamforming-enabled antenna arrays, cognitive spectrum usage, as well as adaptive modulation and coding are a few of the transceiver aspects that are currently tunable in order to optimize the communication efficiency. However, in this optimization process, the wireless environment remains an unmanageable factor; it remains unaware of the communication process undergoing within it \cite{ChritoLI2018,Akyildiz2018mag,husha_LIS1,Kaina_metasurfaces_2014,LIS_twc2018,qignqingwu2019,tang2019wireless,Marco2019}. Furthermore, the wireless environment has in general a harmful effect on the efficiency of wireless links. The signal attenuation limits the connectivity radius of nodes, while multipath propagation resulting in fading phenomena is a well-studied physical factor introducing drastic fluctuations in the received signal power. The signal deterioration is perhaps one of the major concerns in millimeter wave and in the forthcoming TeraHertz (THz) communications \cite{Akyildiz2018mag}.

Although massive MIMO, three-Dimensional (3D) beamforming, and their hardware efficient hybrid analog and digital counterparts \cite{beamforming} provide remarkable approaches to conquer signal attenuation due to wireless propagation via software-based control of the directivity of transmissions, they impose mobility and hardware scalability issues. More importantly, the intelligent manipulation of the ElectroMagnetic (EM) propagation is only partially feasible since the objects in the deployment area, other than the transceivers, are passive and uncontrollable. As a result, the wireless environment as a whole remains unaware of the ongoing communications within it, and the channel model continues to be treated as a probabilistic process, rather than a nearly deterministic one enabled through software-controlled techniques.

Following the recent breakthrough on the fabrication of programmable metamaterials, reconfigurable intelligent surfaces have the potential to fulfill the challenging vision for 6G networks, and materialize seamless connections and intelligent software-based control of the environment in wireless communication systems when coated on the otherwise passive surfaces of various objects \cite{ChritoLI2018,Kaina_metasurfaces_2014,LIS_twc2018,qignqingwu2019}. 
Holographic MIMO Surfaces (HMIMOS) aim at going beyond massive MIMO, being based on low cost, size, weight, and low power consumption hardware architectures that provide a transformative means of the wireless environment into a programmable smart entity \cite{tang2019wireless,qignqingwu2019,LIS_twc2018, Marco2019,han2019,Alkhateeb2019}. In this article, we overview the different emerging HMIMOS architectures and their core functionalities, and discuss their currently considered communication applications as well as their future networking challenges.

\section{HMIMOS Design Models}
In this section, we present available hardware architectures, fabrication methodologies, and operation modes of HMIMOS systems that render them a flexibly integrable concept for diverse wireless communication applications.
\begin{figure*}[!htb] \vspace{-0cm}
            \centering
            \includegraphics[width=17.7cm]{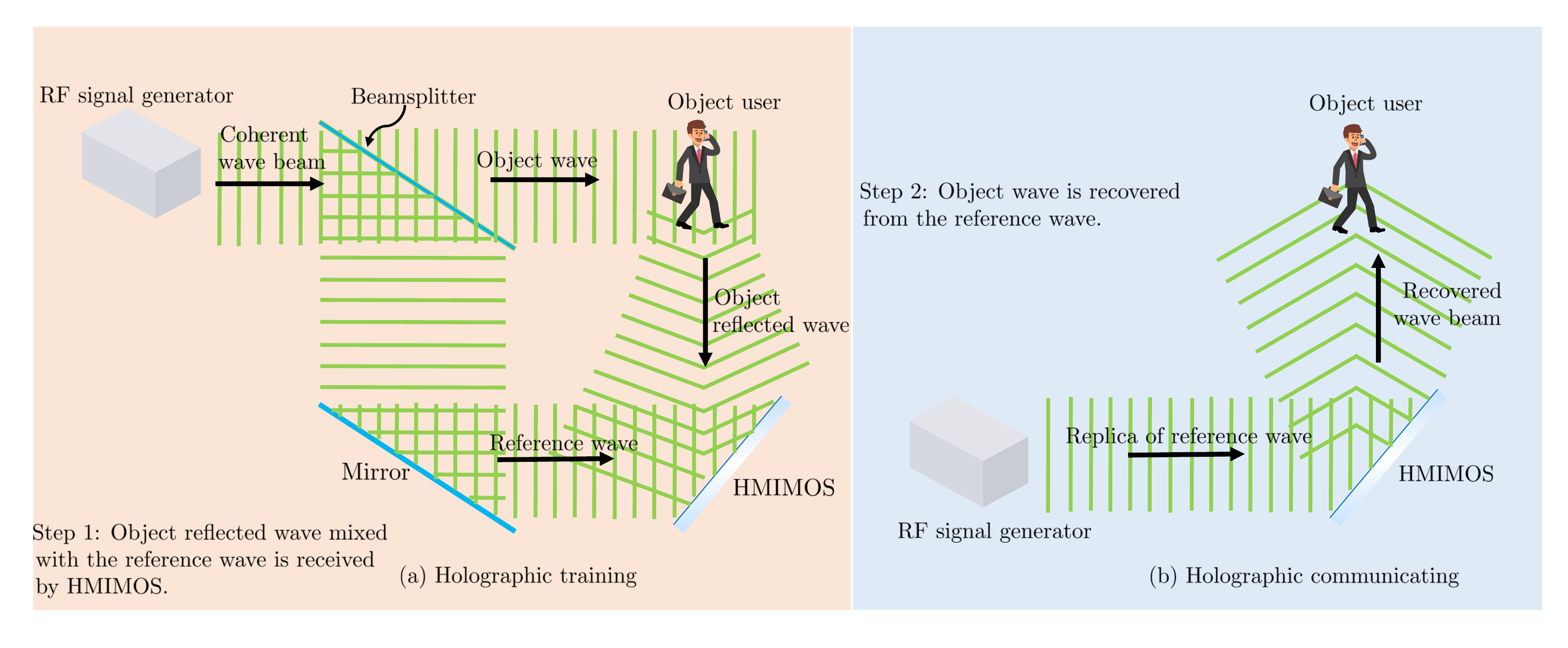} \vspace{-0.2cm}
            \caption{
						The two generic steps of holographic training and holographic communication \cite{holobeamforming}.}  
            \label{fig:hologram} \vspace{-0.5cm}
\end{figure*}\vspace{-0.3cm}

\subsection{Categorization based on the Power Consumption}
\subsubsection{Active HMIMOS}
To realize reconfigurable wireless environments, HMIMOS can serve as a transmitter, receiver, or reflector. When the transceiver role is considered, and thus energy-intensive Radio Frequency (RF) circuits and signal processing units are embedded in the surface, the term active HMIMOS is adopted\cite{holobeamforming,Marzetta2019}. On another note, active HMIMOS systems comprise a natural evolution of conventional massive MIMO systems, by packing more and more software-controlled antenna elements onto a two-Dimensional (2D) surface of finite size. In \cite{husha_LIS1}, where the spacing between adjacent surface elements reduces when their number increase, an active HMIMOS is also termed as Large Intelligent Surface (LIS). A practical implementation of active HMIMOS can be a compact integration of an infinite number of tiny antenna elements with reconfigurable processing networks realizing a continuous antenna aperture. This structure can be used to transmit and receive communication signals across the entire surface by leveraging the hologram principle \cite{holobeamforming,Marzetta2019}. Another active HMIMOS implementation is based on  discrete photonic antenna arrays that integrate active optical-electrical detectors, converters, and modulators for performing transmission, reception, and conversion of optical or RF signals \cite{holobeamforming}.
\subsubsection{Passive HMIMOS}
Passive HMIMOS, also known as Reconfigurable Intelligent Surface (RIS) \cite{ChritoLI2018,Kaina_metasurfaces_2014,LIS_twc2018,tang2019wireless,Marco2019}, or Intelligent Reflecting Surface (IRS) \cite{qingqing2019maga,qignqingwu2019}, acts like a passive metal mirror or `wave collector,' and can be programmed to change an impinging EM field in a customizable way\cite{LIS_twc2018,Marco2019}. Compared with its active counterpart, a passive HMIMOS is usually composed of low cost passive elements that do not require dedicated power sources. Their circuitry and embedded sensors can be powered with energy harvesting modules, an approach that has the potential of making them truly energy neutral. Regardless of their specific implementations, what makes the passive HMIMOS technology attractive from an energy consumption standpoint, is their capability to shape radio waves impinging upon them and forwarding the incoming signal without employing any power amplifier nor RF chain, and also without applying sophisticated signal processing. Moreover, passive HMIMOS can work in full duplex mode without significant self interference or increased noise level, and require only low rate control link or backhaul connections. Finally, passive HMIMOS structures can be easily integrated into the wireless communication environment, since their extremely low power consumption and hardware costs allow them to be deployed into building facades, room and factory ceilings, laptop cases, or even human clothing \cite{LIS_twc2018,Marco2019}.
\begin{figure*}[ht] \vspace{-0.0cm} \hspace{-0.0cm}
            \centering
            \includegraphics[width=17.9cm]{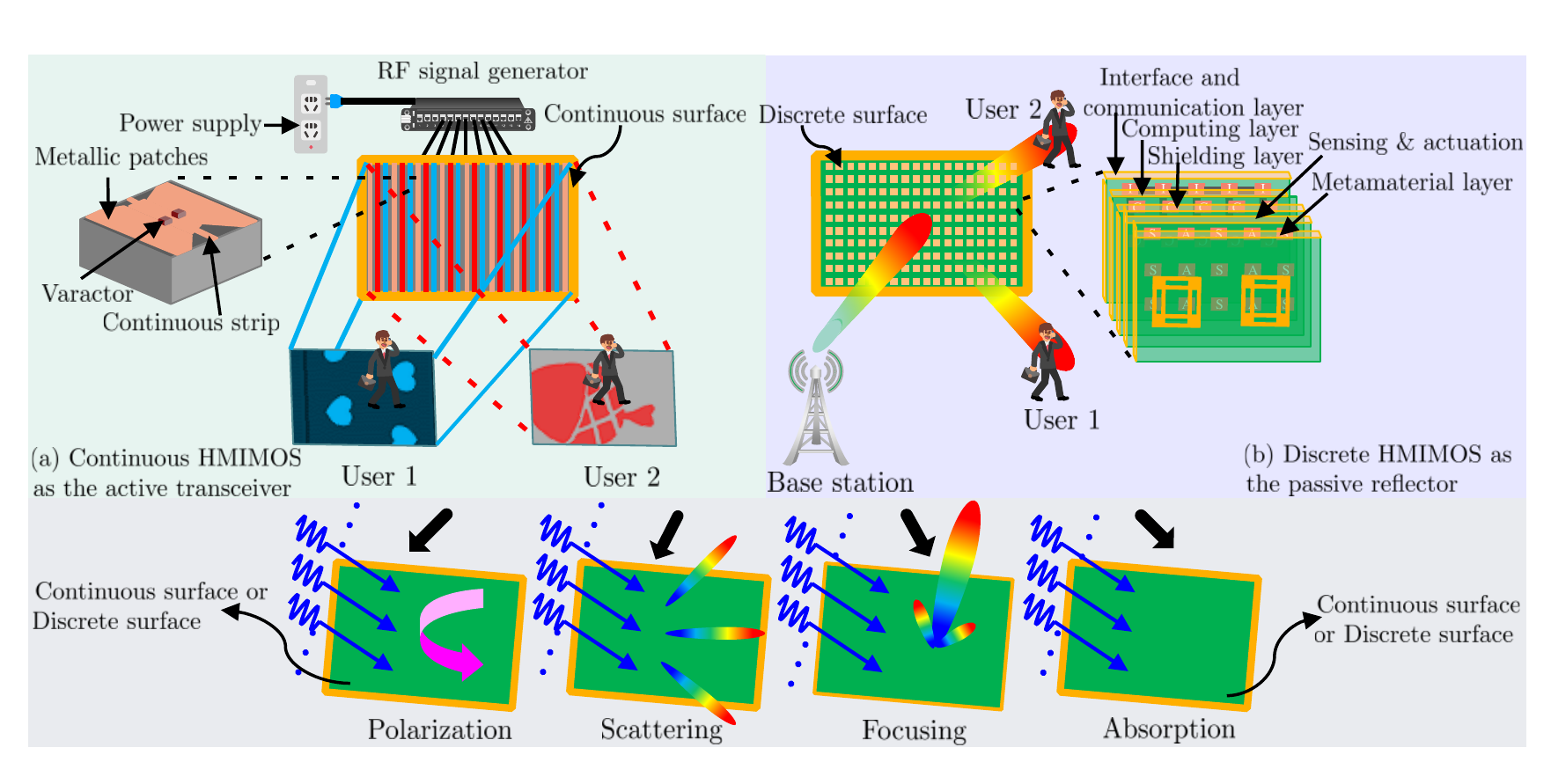} \vspace{-0.0cm}
            \caption{The two operation modes of HMIMOS systems along with their implementation and hardware structures. A schematic view of the HMIMOS functions of EM field polarization, scattering, focusing, and absorption control is provided. }
            \label{fig:examples}\vspace{-0.0cm}
\end{figure*}\vspace{-0.0cm} \vspace{-0.0cm}
\subsection{Categorization based on the Hardware Structure}
\subsubsection{Contiguous HMIMOS}
A contiguous HMIMOS integrates a virtually uncountably infinite number of elements into a limited surface area in order to form a spatially continuous transceiver aperture \cite{holobeamforming,Marzetta2019}. For the better understanding of the operation of contiguous surfaces and their communication models, we commence with a brief description of the physical operation of the optical holography concept. Holography is a technique that enables an EM field, which is generally the result of a signal source scattered off objects, to be recorded based on the interference principle of the EM wave. The recorded EM field can be then utilized for reconstructing the initial field based on the diffraction principle. It should be noted that wireless communications over a continuous aperture is inspired by the optical holography, which is sketched in Fig.~\ref{fig:hologram}. In the training phase, the generated training signals from an RF source are split via a beamsplitter into two waves, the object and reference waves. The object wave is directed to the object, and some of the reflected wave mixed together with the reference wave beam that does not impinge on the object, are fed to the HMIMOS. In the communication phase, the transmitted signal is transformed into the desired beam to the object user over the spatially continuous aperture of the HMIMOS. Since the continuous aperture benefits from the integrated infinite number of antennas that is the asymptotic limit of Massive MIMO, its potential advantages are to achieve higher spatial resolution, and  enable the creation and detection of EM waves with arbitrary spatial frequency components, without undesired side lobes. 

\subsubsection{Discrete HMIMOS}
The discrete HMIMOS is usually composed of many discrete unit cells made of low power software-tunable metamaterials. The means to electronically modify the EM properties of the unit cells range from off the shelves electronic components to using liquid crystals, microelectromechanical systems or even electromechanical switches, and other reconfigurable metamaterials. This structure is substantially different from the conventional MIMO antenna array. One embodiment of a discrete surface is based on discrete `meta-atoms' with electronically steerable reflection properties \cite{ChritoLI2018}. As mentioned earlier, another type of discrete surface is the active one based on photonic antenna arrays. Compared with contiguous HMIMOS, discrete HMIMOS have some essential differences from the perspectives of implementation and hardware, as will be described in the sequel.
\subsection{Fabrication Methodologies}
There are various fabrication techniques for HMIMOS including electron beam lithography at optical frequencies, focused-ion beam milling, interference and nanoimprint lithography, as well as direct laser writing or printed circuit board processes at microwaves. Usually, these fabrication techniques will be ascribed to produce two typical apertures, continuous or discrete apertures, as shown in Fig.~\ref{fig:examples}. A fabrication approach leveraging programmable metamaterials for approximately realizing a continuous microwave aperture \cite{holobeamforming,Marzetta2019} is depicted in Fig.~\ref{fig:examples}(a). This meta-particle structure uses the varactor loading technique to broaden its frequency response range, and achieves continuous aperture and controllable reflection phase. It is a continuous monolayer metallic structure, and comprises an infinite number of meta-particles. Each meta-particle contains two metallic trapezoid patches, a central continuous strip, and varactor diodes. By independently and continuously controlling the bias voltage of the varactors, the surface impedance of continuous HMIMOS can be dynamically programmed, and thus manipulate the reflection phase, amplitude states, and the phase distribution at a wide range of frequency bands \cite{Akyildiz2018mag}. It should be highlighted that this impedance pattern is a map of the hologram, and can be calculated directly from the field distribution of the provided reference wave and reflected object wave, as discussed in Fig.~\ref{fig:hologram}. Exploiting intelligent control algorithms, beamforming can be accomplished by using the hologram principle.

In contrast to continuous aperture, another instance of HMIMOS is the discrete aperture that is usually realized with software-defined metasurface antennas. A general logical structure (regardless of its physical characteristics) was proposed in \cite{ChritoLI2018}, as shown in Fig.~\ref{fig:examples}(b). Its general unit cell structure contains a metamaterial layer, sensing and actuation layers, shielding layer, computing layer, as well as an interface and communications layer with different objectives. Specifically, the meta-material layer is implemented by graphene materials for delivering the desired EM behavior through a reconfigurable pattern, while the objective of sensing and actuation layer is to modify the behavior of the meta-material layer. The shielding layer is made of a simple metallic layer for decoupling the EM behavior of the top and bottom layers to avoid mutual interferences. The computing layer is used to execute external commands from the interface layer or sensors. Finally, the interface and communications layer aims at coordinating the actions of the computing layer and updating other external wireless entities via the reconfigurable interface.

While HMIMOS is in its infancy, basic prototyping work on different kinds of this technology is going on in the world. A discrete HMIMOS was developed by the start-up company named ``Greenerwave'', and which shows the basic feasibility and effectiveness of the HMIMOS concept using the discrete metasurface antennas. In contrast, another start-up company called ``Pivotalcommware'' with the investment of Bill Gates capital is developing the initial commercial products of the contiguous HMIMOS based on the low-cost and contiguous metasurfaces, which further verifies the feasibility of the HMIMOS concept as well as advancement of holographic technologies. Continued  prototyping development is highly desired to both prove the HMIMOS concept with even brand new holographic beamforming technologies and discover potentially new issues that urgently need research. 
\subsection{Operation Modes}
The following four operation modes for HMIMOS are usually considered: 1) continuous HMIMOS as an active transceiver; 2) discrete HMIMOS as a passive reflector; 3) discrete HMIMOS as an active transceiver; and 4) continuous HMIMOS as a passive reflector. Given the recent research interests and due to space limitations, we next elaborate on the first two representative modes of operation, which are also sketched within Fig.~\ref{fig:examples}.
\subsubsection{Continuous HMIMOS as Active Transceivers}
According to this mode of operation, the continuous HMIMOS performs as an active transceiver. The RF signal is generated at its backside and propagates through a steerable distribution network to the contiguous surface constituted by the infinite software-defined and electronically steerable elements that generate multiple beams to the intended users. A distinct difference between active continuous HMIMOS and passively reconfigurable HMIMOS is that the beamforming process of the former is accomplished based on the holographic concept, which is a new dynamic beamforming technique based on software-defined antennas with low cost/weight, compact size, and a low-power hardware architecture.
\subsubsection{Discrete HMIMOS as Passive Reflectors}
Another operation mode of HMIMOS is the mirror or `wave collector,' where the HMIMOS is considered to be discrete and passive. In this case, the HMIMOS include reconfigurable unit cells, as previously described, which makes their beamforming mode resembling that of conventional beamforming\cite{beamforming}, unlike continuous transceiver HMIMOS systems. It is worth noting that most of the existing works (e.g., \cite{Kaina_metasurfaces_2014,LIS_twc2018,qignqingwu2019}) focus on this HMIMOS operation mode which is simpler to implement and analyze.
\section{Functionality, Characteristics, and Communication Applications}
Different fabrication methods of HMIMOS systems result in a variety of functionalities and characteristics, with most of them being very relevant to late expectations for future 6G wireless systems (e.g., Tbps peak rates). In this section, we highlight the HMIMOS functions and key characteristics, and discuss their diverse wireless communications applications.
\subsection{Functionality Types}
Intelligent surfaces can support a wide range of EM interactions, termed hereinafter as functions. Ascribing to their programmable features and depending on whether they are realized via structures with discrete or continuous elements, HMIMOS have four common function types as illustrated in the bottom part of Fig.~\ref{fig:examples}:
\begin{itemize}
  \item \textbf{F1: EM Field Polarization}, which refers to the reconfigurable setting of the oscillation orientation of the wave's electric and magnetic fields.
  \item\textbf{F2: EM Field Scattering}, where the surface redirects an impinging wave with a given direction of arrival towards a desired or multiple concurrent desired directions.
  \item \textbf{F3: Pencile-like Focusing}, which takes place when a HMIMOS acts as lens to focus an EM wave to a given point in the near or far field. The collimation (i.e., the reverse functionality) also belongs in this general mode of beamforming operation.
  \item \textbf{F4: EM Field Absorption}, which implements minimal reflected and/or refracted power of the incoming EM field.
\end{itemize}
\subsection{Characteristics}
Compared with currently used technologies in wireless networks, the most distinctive characteristics of the HMIMOS concept lie in making the environment controllable by providing the possibility of fully shaping and controlling the EM response of the environmental objects that are distributed throughout the network. An HMIMOS structure is usually intended to perform as a signal source or `wave collector' with reconfigurable characteristics, especially for application scenarios where it is used as a passive reflector with the objective of improving the EE. The fundamental properties of HMIMOS systems\footnote{It should be noted that not all HMIMOS architectures have all listed  attributes. Few of them are inherent to passive HMIMOS, but not to active ones, and vice versa. However, we discuss HMIMOS properties here in a broad scope, including all available types up to date.} and their core differences with massive MIMO and conventional multi-antenna relaying systems are summarized as follows:
\begin{itemize}
  \item \textbf{C1: HMIMOS can be nearly passive.} One significant merit of passive HMIMOS is that they do not require any internally dedicated energy source to process the incoming information-carrying EM field.
  \item \textbf{C2: HMIMOS can realize continuous apertures.} Recent research activity focuses on low operational cost methods for realizing spatially-continuous transmitting and receiving apertures.

\item \textbf{C3: Receiver thermal noise is absent in HMIMOS.} Passive HMIMOS do not require to down-convert the received waveform for baseband processing. Instead they implement analog processing directly on the impinging EM field.
\item \textbf{C4: HMIMOS elements are tuned in software.} Available architectures for metasurfaces enable simple reprogrammability of all settings of their unit elements.
  \item \textbf{C5: HMIMOS can have full-band response.} Due to recent advances in meta-materials' fabrication, reconfigurable HMIMOS can operate at any operating frequency, ranging from the acoustic spectrum up to THz and the light spectra.
  \item \textbf{C6: Distinctive low latency implementation.} HMIMOS are based on rapidly  reprogrammable meta-materials, whereas conventional relaying and massive MIMO systems include antenna array architectures.
\end{itemize}

\subsection{Communications Applications}
\begin{figure*}[!htb] \vspace{-0.5cm}
            \centering
            \includegraphics[width=16cm]{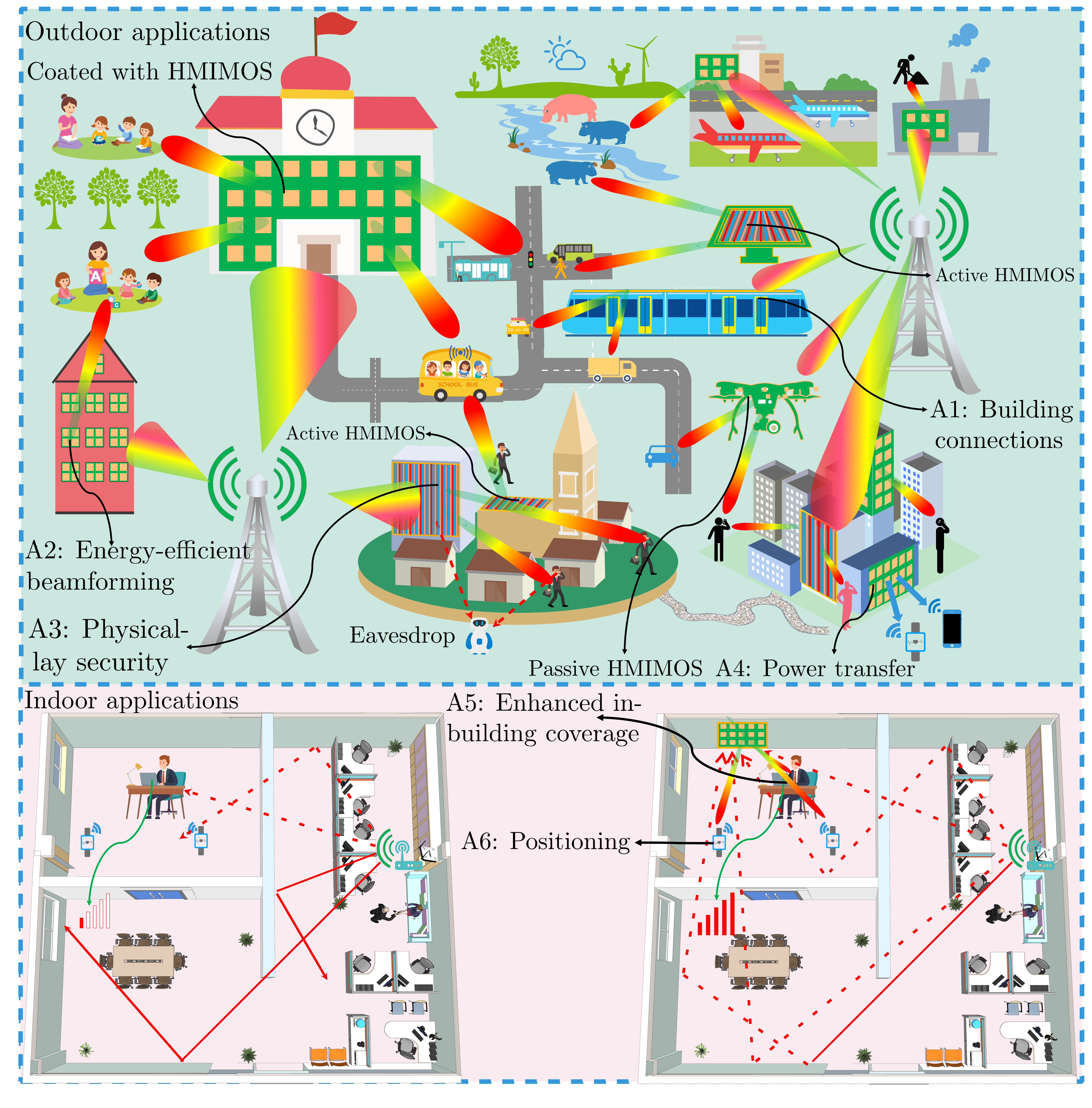} \vspace{-0.2cm}
            \caption{Wireless communications applications of HMIMOS in outdoor and indoor environments.}  \vspace{-0.4cm}
            \label{fig:application}
\end{figure*}\vspace{-0.0cm} \vspace{-0.0cm}
The unique features of HMIMOS enabling intelligent and rapidly reconfigurable wireless environments make them an emerging candidate technology for the low-power, high-throughput, and low-latency vision of 6G wireless networks. We next discuss representative communications applications of HMIMOS for outdoor and indoor environments.

\subsubsection{Outdoor Applications}
Consider the discrete passive HMIMOS as an indicative example that comprises a finite number of unit elements, and intended for forwarding suitably phase-shifted versions of its impinging signals to users over different outdoor scenarios, such as typical urban, shopping malls, and international airports, as illustrated in the upper part of Fig$.$~\ref{fig:application}. We assume that HMIMOS are planar structures of few centimeters thickness and variable sizes that can be easily deployed onto nearly all environmental objects.
\begin{itemize}
  \item \textbf{A1: Building connections.} HMIMOS can extend the coverage from outdoor BSs to indoor users, especially in cases where there is no direct link between the users and BS, or the link is severely blocked by obstacles.
  \item \textbf{A2: Energy-efficient beamforming.} HMIMOS are capable of recycling ambient EM waves and focusing them to their intended users via effective tuning of their unit elements. In such cases, surfaces are deployed as relays to forward the information bearing EM field to desired locations via efficient beamforming that compensates for the signal attenuation from the BS or co-channel interference from neighboring BSs.
	\item \textbf{A3: Physical-layer security.} HMIMOS can be deployed for physical layer security in order to cancel out reflections of the BS signals to eavesdroppers.
  \item \textbf{A4: Wireless power transfer.} HMIMOS can collect ambient EM waves and direct them to power-hungry IoT devices and sensors enabling also simultaneous wireless information and power transfer.
\end{itemize}
\begin{table*}[t]
\caption{Some recent research results on HMIMOS-based wireless communication systems.} \label{features}  \vspace{-1mm}
\begin{center}
    \begin{tabular}{| p{0.9cm} | p{0.1cm} | p{1.4cm} | p{1.2cm}  | p{1.7cm}  | p{8.0cm} |  }
    \hline
\multicolumn{2}{|c|}{\textbf{Related Works}}   &{\textbf{Applications}} &{\textbf{Functions}} & {\textbf{Characteristics}} & \textbf{Main Contributions} \\    \hline

~   & \vspace{-1.8mm} \cite{Akyildiz2018mag} & \vspace{1.5mm} A1,  A2, A5 &  \vspace{1.5mm} F2, F3  & \vspace{1.5mm} C1, C3-C6 & Presented a HMIMOS-based approach to combat the distance limitation in millimeter wave and THz systems; simulation results for an indoor set up corroborated the merits of proposed approach. \\    \cline{2-6}	

\multirow{2}*{Indoor}  & \vspace{-3mm} \cite{husha_LIS1}  & \vspace{0mm} A2, A5, A6  & \vspace{0mm} F2, F3 & \vspace{0mm} C2-C6 &  Introduced an indoor signal propagation model and presented information theoretical results for active and continuous HMIMOS systems.  \\    \cline{2-6}

~   & \vspace{-2mm} \cite{ChritoLI2018} & \vspace{1.5mm} A1-A3, A5 &  \vspace{1.5mm} F1-F4 & \vspace{1.5mm} C1, C3-C6 &  Introduced the concept of programmable indoor wireless environments offering simultaneous communication and security; an indoor model and a simulation set up for HMIMOS communication were presented. \\    \cline{2-6}	

~  & \vspace{-2.5mm} \cite{Kaina_metasurfaces_2014} & \vspace{1mm} A1, A2 & \vspace{1mm} F2, F3  & \vspace{1mm} C1, C3-C6 &  Designed a 0.4m$^2$ and 1.5mm thick planar metasurface consisting of $102$ controllable unit cells operating at $2.45$GHz; demonstrated increased received signal strength when deployed indoors. \\    \cline{2-6}

~  & \vspace{-2.5mm} \cite{tang2019wireless} & \vspace{1mm} A1, A2, A5 &  \vspace{1mm} F2, F3  & \vspace{1mm} C3-C6 & Proposed free space pathloss models using the EM and physical properties of a reconfigurable surface; indoor field experiments validated the proposed models.   \\    \hline

~  & \vspace{-2.5mm} \cite{LIS_twc2018} & \vspace{1mm} A1, A2 &   \vspace{1mm} F2, F3 & \vspace{1mm} C1, C3-C6 &  Proposed HMIMOS for outdoor MIMO communications and presented EE maximization algorithms; studied the fundamental differences between HMIMOS and conventional multi-antenna relays.   \\    \cline{2-6}

~ & \vspace{-2.5mm} \cite{qignqingwu2019}  & \vspace{1mm} A2, A4 &  \vspace{1mm} F2, F3  & \vspace{1mm} C1, C3-C6 &  Presented jointly active and passive beamforming algorithms for HMIMOS-assisted MIMO communication; analyzed the interference distribution and studied the power scaling law. \\    \cline{2-6}



\multirow{2}*{Outdoor}  & \vspace{-2.5mm} \cite{han2019}   & \vspace{1mm} A2 & \vspace{1mm} F2, F3 & \vspace{1mm} C1, C3-C6 &  Derived the optimal HMIMOS phase matrix for the case of available statistical channel information and presented a tight approximation for the ergodic capacity. \\    \cline{2-6}


~  & \vspace{-4mm} \cite{Alkhateeb2019} & \vspace{0mm} A1, A2 & \vspace{0mm} F1-F4 &  \vspace{0mm} C1, C3-C6 &  Studied compressive sensing and deep learning approaches for HMIMOS channel estimation and online configuration. \\    \hline

    \end{tabular}
    \label{table:features}
\end{center}  \vspace{-3mm}
\end{table*} \vspace{-2mm}
\subsubsection{Indoor Applications}
Indoor wireless communication is subject to rich multipath propagation due to the presence of multiple scatters and signal blocking by walls and furniture, as well as RF pollution due to the highly probable densification of electronic devices in confined spaces. As such, providing ubiquitous high throughput indoor coverage and localization is a challenging task. HMIMOS has the potential of being highly beneficial in indoor environments, leveraging from its inherit capability to reconfigure EM waves towards various communication objectives. An illustrative general example is sketched in the lower part of Fig.~\ref{fig:application}. In the left corner of this example where a HMIMOS is absent, the signal experiences pathloss and multipath fading due to refraction, reflection, and diffusion, which deteriorates its sufficient propagation to the target user. However, in the right corner of Fig.~\ref{fig:application}, signal propagation can be boosted using HMIMOS coated in the wall so as to assist the signal from the access point to reach the intended user with the desired power level.
\begin{itemize}
  \item \textbf{A5: Enhanced in-building coverage:} As previously discussed, indoor environments can be coated with HMIMOS to increase the throughput offered by conventional Wi-Fi access points.
  \item \textbf{A6: High accurate indoor positioning}: HMIMOS has increased potential for indoor positioning and localization, where the conventional Global Positioning System (GPS) fails. Large surfaces offer large, and possibly continuous, apertures that enable increased spatial resolution.
\end{itemize}

There has been lately increasing research interest in wireless communication systems incorporating HMIMOS. In Table~\ref{table:features}, we list some of the recent works dealing with different combinations among the functionalities of HMIMOS, their characteristics, and communication applications.

\section{Design Challenges and Opportunities}
In this section, we present some main theoretical and practical challenges with HMIMOS-based communication systems. \vspace{-0.2cm}
\subsection{Fundamental Limits}
It is natural to expect that wireless communication systems incorporating HMIMOS will exhibit different features compared with traditional communications based on conventional multi-antenna transceivers. Recall that current communication systems operate over uncontrollable wireless environments, whereas HMIMOS-based systems will be capable of reconfiguring their EM propagation impact. This fact witnesses the need for new mathematical methodologies to characterize the physical channels in HMIMOS-based systems and analyze their ultimate capacity gains over a given volume \cite{Marzetta2019}, as well as for new signal processing algorithms and networking schemes for realizing HMIMOS-assisted communication. For example, continuous HMIMOS is used for the reception and transmission of the impinging EM field over its continuous aperture using the hologram concept. Differently from the currently considered massive MIMO systems, HMIMOS operation can be described by the Fresnel-Kirchhoff integral that is based on the Huygens-Fresnel principle \cite{holobeamforming}.
\begin{figure*}[ht]
\centering
\subfloat[]{
\label{fig:indoorp}
\begin{minipage}[c]{0.3\textwidth}
\centering
\centerline{\includegraphics[width=9cm]{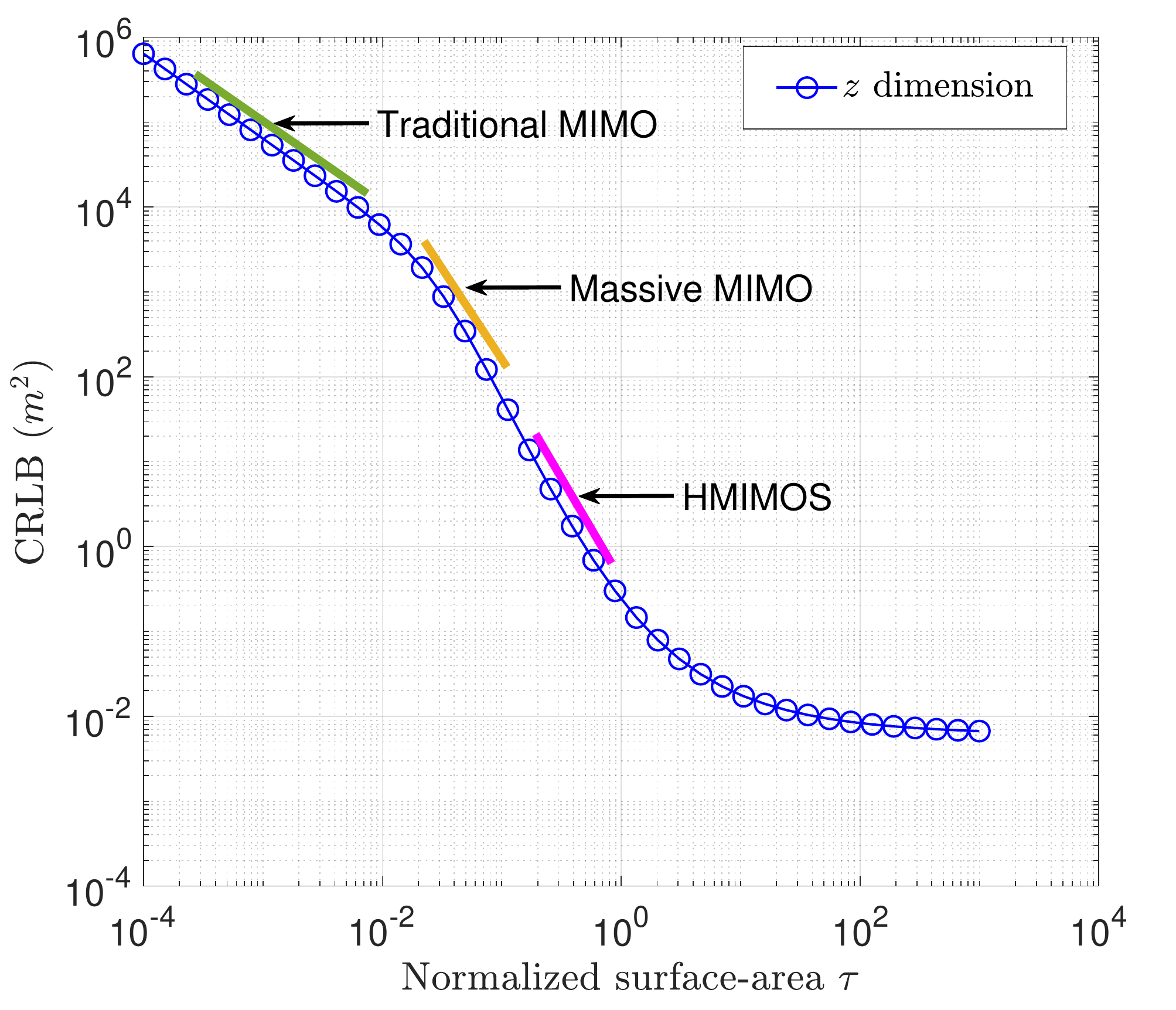}}
\end{minipage}
} \hspace{35mm}
\centering
\subfloat[]{
\label{fig:ee}
\begin{minipage}[c]{0.3\textwidth}
\centering
\centerline{\includegraphics[width=9cm]{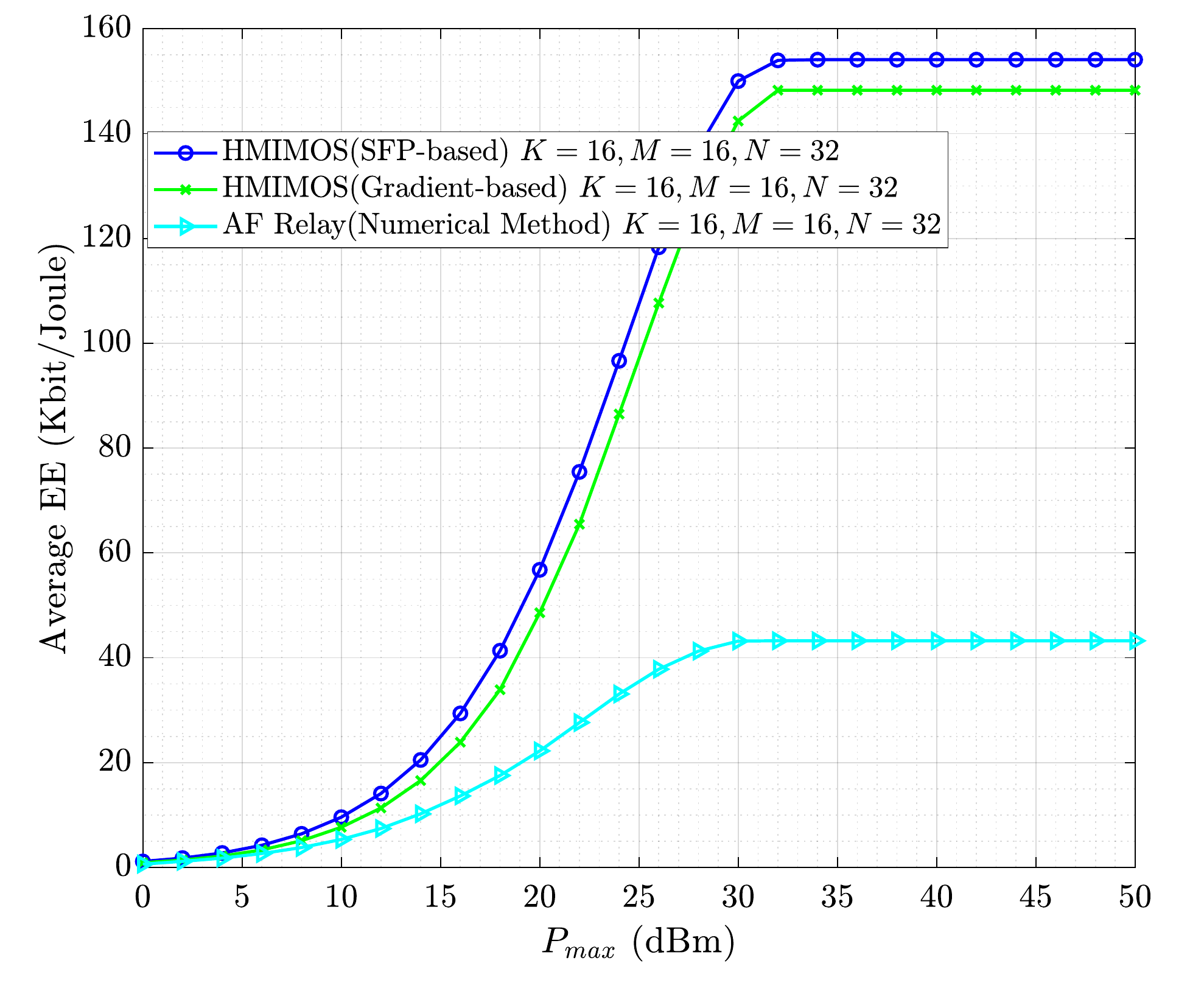}}
\end{minipage}
}
\caption{
(a)	CRLBs of positioning with an active HMIMOS of a radius $R$ for the case where a single user is located $z=4m$ away from the center of surface. The wavelength $\lambda$ is $0.1m$, and $\tau$ represents the normalized surface-area~\cite{husha_LIS1}. (b) Average EE with HMIMOS-assisted communication versus the maximum BS transmit power $P_{\rm max}$ in dB. }
\label{fig:ee_all}
\end{figure*} 
\subsection{HMIMOS Channel Estimation}
The estimation of possibly very large MIMO channels in HMIMOS-based communication systems is another critical challenge due to the various constraints accompanying the available HMIMOS hardware architectures. Most of the few currently available approaches mainly consider large time periods for training all HMIMOS unit elements via pilots sent from the BS and received at the user equipment via generic reflection. Another family of techniques employs compressive sensing and deep learning via online beam/reflection training for channel estimation and design of the phase matrix\cite{Alkhateeb2019}. However, this mode of operation requires large amounts of training data, and employs fully digital or hybrid analog and digital transceiver architectures for HMIMOS, which will results in increased hardware complexity and operational power consumption.

\subsection{Robust Channel-Aware Beamforming}
Channel dependent beamforming has been extensively considered in massive MIMO systems. However, realizing environment-aware designs in HMIMOS-based communication systems is extremely challenging, since the HMIMOS unit cells that are fabricated from metamaterials impose demanding tuning constraints. The latest HMIMOS design formulations (e.g., \cite{LIS_twc2018,qignqingwu2019}) include large numbers of reconfigurable parameters with non-convex constraints rendering their optimal solution highly non-trivial. For the case of continuous HMIMOS, intelligent holographic beamforming is the objective in order to smartly target and follow individual or small clusters of devices, and provide them with high-fidelity beams and smart radio management. However, self-optimizing holographic beamforming technologies that depend on complex aperture synthesis and low level modulation are not available yet.

\subsection{Distributed Configuration and Resource Allocation}
Consider a HMIMOS-based communication system comprising multiple multi-antenna BSs, multiple HMIMOS, and massive number of users, where each user is equipped with a single or multiple antennas. The centralized configuration of HMIMOS will require massive amounts of control information to be communicated to a central controller, which is prohibitive both in terms of computational overhead and energy consumption. Hence, distributed algorithms for the EE-optimal resource allocation and beamforming, HMIMOS configurations, and users' scheduling need to be developed. Additional optimization parameters complicating the network optimization are anticipated to be the power allocation and spectrum usage, as well as the users' assignment to BSs and distributed HMIMOS. Naturally, the more HMIMOS are incorporated in the network, the more challenging the algorithmic design will becomes.

\section{Case studies}
In this section, we summarize the performance of HMIMOS in two typical application scenarios: indoor positioning with an active continuous HMIMOS and outdoor downlink communication assisted by a passive discrete HMIMOS. \vspace{-0.0cm}
\begin{table*}  
\caption{Simulation Parameters for the Average EE Performance Results in Fig.~\ref{fig:ee}.} \label{tabpar} \vspace{-0mm}
\begin{center}
    \begin{tabular}{| l | l || l | l |}
    \hline
    \textbf{Parameters} & \textbf{Values} & \textbf{Parameters} & \textbf{Values} \\ \hline
    HMIMOS central element placement: & $(100m,100m)$ & Circuit dissipated power at BS: & $9$dBW  \\
    BS central element placement: & $(0m,0m)$ &  Circuit dissipated power coefficients at BS and AF relay: & $1.2$  \\
		Small scale fading model: & Rayleigh &     Maximum transmit power at BS and AF relay $P_{\rm max}$: & $20$dBW \\
		Large scale fading model at distance $d$: & $10^{-3.53}d^{-3.76}$ & Dissipated power at each user: & $10$dBm  \\
    Transmission bandwidth: & $180\mathrm{KHz}$  & Dissipated power at each HMIMOS element: & $10$dBm  \\
    Algorithmic convergence parameter: & $\epsilon=10^{-3}$  &      Dissipated power at each AF relay transmit-receive antenna: & $10$dBm \\
    \hline
    \end{tabular}
\end{center} 
\end{table*}
\subsection{Indoor Positioning with an Active Continuous HMIMOS}
We assume an active HMIMOS where the distance between any of each two adjacent unit elements is $\lambda/2$, with $\lambda$ being the carrier wavelength. In such a discretized manner, traditional MIMO, massive MIMO, and HMIMOS are unified, and the differences lie in the number of antenna elements used, i.e., the surface area. It was shown in \cite{husha_LIS1} that the number of antennas in a traditional massive MIMO system for a given surface area $\pi R^2$ is equal to $\pi R^2\!/(\lambda^2/4)\!=\!\pi \tau z^2/(\lambda^2/4)\cong20106\tau$, when $z\!=\!4m$, $\lambda\!=\!0.1m$, and $\tau\!\triangleq\!(R/z)^2$ (the normalized surface area). A typical massive MIMO array comprising of $N\!=\!200$ antennas results in $\tau\!\approx\!0.01$, while an active HMIMOS typically increases the surface-area (so as $\tau$) by $10\!\sim\!20$ \cite{husha_LIS1}. In Fig.~\ref{fig:indoorp}, the Cram\'{e}r--Rao Lower Bounds (CLRBs) of user positioning in the presence of phase uncertainty are sketched. As depicted, the CRLB of positioning decreases linearly with $\tau$ for traditional MIMO, while massive MIMO falls short in reaching the cubic decreasing slope that is achieved by the active HMIMOS, yielding significant gains in user positioning.

\subsection{EE Maximization with a Passive Discrete HMIMOS}
We consider an outdoor $16$-antenna BS simultaneously serving $16$ single-antenna users in the downlink communication using a discrete passive HMIMOS with $32$ unit elements that is attached to a surrounding building's facade \cite{LIS_twc2018}, and the simulation parameters are shown in Table~\ref{tabpar} \cite{LIS_twc2018}. The obtained EE performance using an approach based on Sequential Fractional Programming (SFP), as well as a gradient descent approach, to tune the HMIMOS system is sketched in Fig$.$~\ref{fig:ee} as a function of the maximum BS transmit power $P_{\rm max}$. We have also numerically evaluated the EE of conventional Amplify-and-Forward (AF) relaying. It is shown that the HMIMOS-assisted system achieves a three-fold increase in EE compared to the AF relaying case when $P_{\rm max}\geq32$dBm. When $P_{\rm max}\geq32$dBm, EE performance saturates, which reveals that the excess BS transmit power should not be used because it would decrease EE.
\section{Conclusion}
In this article, we investigated the emerging concept of HMIMOS wireless communication, and in particular the available HMIMOS hardware architectures, their functionalities and characteristics, as well as their recent communication applications. We highlighted their great potential as a key enabling technology for the physical layer of future 6G wireless networks. HMIMOS technology offers huge advantages in terms of SE and EE, yielding smart and reconfigurable wireless environments. It reduces the cost, size, and energy consumption of network devices, providing ubiquitous coverage and intelligent communication in both indoor and outdoor scenarios. Benefiting from its merits, HMIMOS can be compactly and easily integrated into a wide variety of applications. Representative use cases are the extension of coverage, physical-layer security, wireless power transfer, and positioning. However, there are still challenges ahead to achieve the full potential of this emerging technology. Among them belong the realistic modeling of metasurfaces, the analysis of the fundamental limits of wireless communications with multiple HMIMOS, the implementation of intelligent environment-aware adaptation, and the channel estimation with nearly passive surfaces. These challenges provide academic and industrial researchers with a gold mine of new problems and challenges to tackle.

\bibliographystyle{IEEEbib}
\bibliography{reference_mag}

\end{document}